\documentclass[12pt]{article}
\usepackage{amsmath,amssymb}
\newcommand{\eqdef}{\stackrel{\text{def}}{=}}

\allowdisplaybreaks[3]

\begin{document}

\baselineskip=20pt

%%%%%%%%%%%%%%%%%%%%%%%%%%%%%%%%%%%%%%%%%%%%%%%%%%%%%%%%%%%%
%                                                          %
%  Title page                                              %
%                                                          %
%%%%%%%%%%%%%%%%%%%%%%%%%%%%%%%%%%%%%%%%%%%%%%%%%%%%%%%%%%%%
\newfont{\elevenmib}{cmmib10 scaled\magstep1}
\newcommand{\preprint}{
     \begin{flushleft}
       \elevenmib Yukawa\, Institute\, Kyoto\\
     \end{flushleft}\vspace{-1.3cm}
     \begin{flushright}\normalsize  \sf
       DPSU-07-4\\
       YITP-07-64\\
      {\tt arXiv:0710.2209[hep-th]}\\
       October 2007
     \end{flushright}}
\newcommand{\Title}[1]{{\baselineskip=26pt
     \begin{center} \Large \bf #1 \\ \ \\ \end{center}}}
\newcommand{\Author}{\begin{center}
     \large \bf Satoru Odake${}^a$ and Ryu Sasaki${}^b$ \end{center}}
\newcommand{\Address}{\begin{center}
       $^a$ Department of Physics, Shinshu University,\\
       Matsumoto 390-8621, Japan\\
       ${}^b$ Yukawa Institute for Theoretical Physics,\\
       Kyoto University, Kyoto 606-8502, Japan
     \end{center}}

\preprint
\bigskip\bigskip\bigskip

\Title{$q$-oscillator from the $q$-Hermite Polynomial}

\Author

\Address

\begin{abstract}
By factorization of the Hamiltonian describing the quantum mechanics
of the continuous $q$-Hermite polynomial, the creation and annihilation
operators of the $q$-oscillator are obtained.
They satisfy a $q$-oscillator algebra as a consequence of the
shape-invariance of the Hamiltonian. A second set of $q$-oscillator is
derived from the exact Heisenberg operator solution.
Now the $q$-oscillator stands on the equal footing to the ordinary
harmonic oscillator.
\end{abstract}

PACS : 03.65.-w, 03.65.Ca, 03.65.Fd, 02.30.Ik, 02.30.Gp, 02.20.Uw

%%%%%%%%%%%%%%%%%%%%%%%%%%%%%%%%%%%%%%%%%%%
% PACS 2006
% 02.       Mathematical methods in physics
%   02.20.Uw  Quantum Groups
%   02.30.Gp  Special functions
%   02.30.Ik  Integrable systems
% 03.       Quantum mechanics, field theories, and special relativity
%   03.65.-w  Quantum mechanics
%   03.65.Ca  Formalism
%   03.65.Fd  Algebraic methods
%   03.65.Ge  Solutions of wave equations: bound states
% 45.       Classical mechanics of discrete systems
%   45.20.-d  Formalisms in classical mechanics
%%%%%%%%%%%%%%%%%%%%%%%%%%%%%%%%%%%%%%%%%%%

%%%%%%%%%%%%%%%%%%%%%%%%%%%%%%%%%%%%%%%%%%%
% PACS 2008
% 02.00.00 Mathematical methods in physics
%   02.20.Uw Quantum groups
%   02.30.Gp Special functions
%   02.30.Ik Integrable systems
% 03.00.00 Quantum mechanics, field theories, and special relativity
%   03.65.-w Quantum mechanics
%   03.65.Ca Formalism
%   03.65.Db Functional analytical methods
%   03.65.Fd Algebraic methods
%   03.65.Ge Solutions of wave equations: bound states
% 45.00.00 Classical mechanics of discrete systems
%   45.20.-d Formalisms in classical mechanics
%%%%%%%%%%%%%%%%%%%%%%%%%%%%%%%%%%%%%%%%%%%

%\keywords{quantum. etc}

%\preprint{{\texttt{arXiv:yymm.nnnn[hep-th??]}},
% {\textsf{DPSU-07-4}}, {\textsf{YITP-07-??}}}

%\maketitle

%\newpage
%%%%%%%%%%%%%%%%%%%%%%%%%%%%%%%%%%%%%%%%%%%%%%%%%%%%%%%%%%%%%%%
%                                                             %
%  1. Introduction                                            %
%                                                             %
%%%%%%%%%%%%%%%%%%%%%%%%%%%%%%%%%%%%%%%%%%%%%%%%%%%%%%%%%%%%%%%
\section{Introduction}
\label{intro}
\setcounter{equation}{0}

In this Letter, the explicit forms of the generators of a $q$-oscillator
algebra are derived from the quantum mechanical Hamiltonian
\cite{os5,os4} of the $q$-Hermite polynomial \cite{rogers}, the
$q$-analogue of the Hermite polynomial constituting the eigenfunctions
of the harmonic oscillator.
This is in sharp contrast to the common approach to $q$-oscillators
\cite{qoscillators}, which assumes certain forms of the algebras without
any dynamical/analytical contents behind them.
On the other hand, the ordinary harmonic oscillator algebra generated
by the annihilation/creation operators  has rich analytical structure of
differential operators related with the classical analysis of the
Hermite polynomial together with the coherent and squeezed states, etc.
Since the annihilation/creation operators of the harmonic oscillator and
their algebra are the cornerstone of modern quantum physics, their good
deformation is bound to play an important role, as evidenced by the
representation theory of the quantum groups in terms of the
$q$-oscillators.
Thus our new results are expected to enrich the subject by stimulating
the interplay between (quantum) algebra and analysis through new
coherent/squeezed states etc, which would find applications in quantum
optics and quantum information theory.
Here we discuss only Rogers' $q$-Hermite polynomial \cite{rogers}, or
the so-called continuous $q$-Hermite polynomial \cite{koeswart,andaskroy} 
for the parameter range $0<q<1$. Like the Hermite polynomial,
the $q$-Hermite polynomial has no parameter other than $q$.

This Letter is organized as follows. The factorized Hamiltonian for the
$q$-Hermite polynomial is presented and the $q$-oscillator commutation
relation is shown to be a simple consequence of their structure.
After brief exploration of the eigenfunctions, the exact Heisenberg
operator solution \cite{os7} is presented.
A second set of $q$-oscillator algebra is derived from the explicit
forms of the annihilation/creation operators which are the
positive/negative energy parts of the exact Heisenberg operator solution.
These $q$-oscillators reduce to the ordinary harmonic oscillator in 
the $q\to 1$ limit.
Relationship to various forms of $q$-oscillator algebras is explained.
The Letter concludes with some historical comments and a summary.

%%%%%%%%%%%%%%%%%%%%%%%%%%%%%%%%%%%%%%%%%%%%%%%%%%%%%%%%%%%%%%%
%                                                             %
%  2. Hamiltonian for the $q$-Hermite polynomial              %
%                                                             %
%%%%%%%%%%%%%%%%%%%%%%%%%%%%%%%%%%%%%%%%%%%%%%%%%%%%%%%%%%%%%%%
\section{Hamiltonian for the $q$-Hermite polynomial}
\label{Hams}

The Hamiltonian of the `discrete' quantum mechanics for one degree of
freedom has the general structure \cite{os4,os5}
\begin{align}
   \mathcal{H}&\eqdef\sqrt{V(x)}\,e^{\gamma p}\sqrt{V(x)^*}
   +\sqrt{V(x)^*}\,e^{-\gamma p}\sqrt{V(x)}-V(x)-V(x)^*
   \label{H}\\
   &=\sqrt{V(x)}\,q^{D}\sqrt{V(x)^*}
   +\sqrt{V(x)^*}\,q^{-D}\sqrt{V(x)}-V(x)-V(x)^*,
   \label{H2}
\end{align}
in which $x\in\mathbb{R}$ is the coordinate and $p=-i\partial_x$ is
the conjugate momentum.
The constant $\gamma$ in the present case is $\gamma\eqdef\log q$,
$0<q<1$ and the potential function for the dynamics of the $q$-Hermite
polynomial is given by
\begin{equation}
   V(x)\eqdef \frac{1}{(1-z^2)(1-qz^2)},\quad z\eqdef e^{ix},
   \label{Vform}
\end{equation}
with $D=p=-i\partial_x=z\frac{d}{dz}$.
It is a special case of the Askey-Wilson polynomial \cite{os5,koeswart}.
The Hamiltonian is factorized as
\begin{align}
   \mathcal{H}&=\mathcal{A}^\dagger\mathcal{A},
   \label{factorise}\\
   \mathcal{A}^\dagger&\eqdef
   -i\bigl(\sqrt{V(x)}\,q^{D/2}-\sqrt{V(x)^*}\,q^{-D/2}\bigr),
   \label{A}\\
   \mathcal{A}&\eqdef
   i\bigl(q^{D/2}\sqrt{V(x)^*}-q^{-D/2}\sqrt{V(x)}\,\bigr).
   \label{defA}
\end{align}
With the explicit form of the potential function $V$, \eqref{Vform},
it is straightforward to derive the $q$-oscillator commutation relation
\begin{equation}
   \mathcal{A}\mathcal{A}^{\dagger}-q^{-1}\mathcal{A}^{\dagger}\mathcal{A}
   =q^{-1}-1.
   \label{qrel}
\end{equation}
Sometimes it is written as
$[\mathcal{A}, \mathcal{A}^{\dagger}]_{q^{-1}}=q^{-1}-1$
with the standard notation $[A,B]_c\eqdef AB-cBA$.
We also have
\begin{equation}
   [\mathcal{H},\mathcal{A}]_q=(q-1)\mathcal{A},\quad
   [\mathcal{H},\mathcal{A}^{\dagger}]_{q^{-1}}
   =(q^{-1}-1)\mathcal{A}^{\dagger}.
   \label{qrel2}
\end{equation}
The $q$-oscillator commutation relation \eqref{qrel} is also a
consequence of the {\em shape invariance\/} without shifting parameter
\cite{genden} among the general
Askey-Wilson potentials \cite{os5,koeswart}. One could also say that the
commutation relation of the harmonic oscillator
$aa^{\dagger}-a^{\dagger}a=1$ is a manifestation of the shape-invariance.

The groundstate wavefunction $\phi_0$ is annihilated by the operator
$\mathcal{A}$:
\begin{equation}
   \mathcal{A}\phi_0=0\ \Longrightarrow
   \phi_0(x)\eqdef \sqrt{(e^{2ix};q)_\infty(e^{-2ix};q)_\infty},
\end{equation}
in which the standard notation of $q$-Pochhammer symbol $(a\,;q)_n$ is
used:
\begin{equation}
   (a\,;q)_n\eqdef\prod_{k=1}^n(1-aq^{k-1})
   =(1-a)(1-aq)\cdots(1-aq^{n-1}),
   \label{defqPoch}
\end{equation}
including the limiting case $n\to\infty$.
With this choice of the groundstate wavefunction, we can show that
the Hamiltonian \eqref{H} is hermitian with respect to the inner
product $(f,g)=\int_0^{\pi}f(x)^*g(x)dx$ in the
Hilbert space $L^2[0,\pi]$ \cite{sasaki07}.
By using the factorization \eqref{factorise} and the $q$-oscillator
relation \eqref{qrel},
it is straightforward to demonstrate that $(\mathcal{A}^{\dagger})^n\phi_0$
is an eigenstate of the Hamiltonian with the geometric sequence spectrum:
\begin{equation}
   \mathcal{H}(\mathcal{A}^{\dagger})^n\phi_0=
   \mathcal{E}_n(\mathcal{A}^{\dagger})^n\phi_0,\quad
   \mathcal{E}_n\eqdef q^{-n}-1.
   \label{spectrum}
\end{equation}

%%%%%%%%%%%%%%%%%%%%%%%%%%%%%%%%%%%%%%%%%%%%%%%%%%%%%%%%%%%%%%%
%                                                             %
%  3. The $q$-Hermite polynomial                              %
%                                                             %
%%%%%%%%%%%%%%%%%%%%%%%%%%%%%%%%%%%%%%%%%%%%%%%%%%%%%%%%%%%%%%%
\section{The $q$-Hermite polynomial}
\label{q-Hermite}

The analytical approach to the Schr\"{o}dinger equation
\begin{equation}
   \mathcal{H}\phi_n=\mathcal{E}_n\phi_n,
\end{equation}
which is a {\em difference\/} equation instead of a second order
differential equation, goes as follows.
By similarity transformation in terms of the groundstate wavefunction
$\phi_0$, one introduces
\begin{equation}
   \widetilde{\mathcal{H}}\eqdef
   \phi_0^{-1}\circ \mathcal{H}\circ \phi_0
   =V(x)(q^D-1)+V(x)^*(q^{-D}-1),
\end{equation}
which acts on the polynomial part of the eigenfunction $P_n(\eta(x))$:
\begin{equation}
   \phi_n(x)=\phi_0(x) P_n(\eta(x)).
\end{equation}
It is elementary to show
\begin{equation}
   \widetilde{\mathcal{H}}\,(z+1/z)^n=
   (q^{-n}-1)(z+1/z)^n
   +\text{lower order terms in }z+1/z,
\end{equation}
since the residues at $z=\pm1$, $z=\pm q^{\pm1/2}$, and
$z=\pm q^{\mp1/2}$ all vanish.
Thus one can find the eigenpolynomial in $\eta(x)=\cos x=(z+1/z)/2$,
which is called the continuous $q$-Hermite polynomial introduced by
Rogers \cite{rogers,koeswart}
\begin{align}
   &\widetilde{\mathcal{H}}\,H_n(\cos x|q)=\mathcal{E}_nH_n(\cos x|q),\\
   &H_n(\cos x|q)\eqdef
%   \sum_{k=0}^n\frac{(q\,;q)_n}{(q\,;q)_k(q\,;q)_{n-k}}\cos[(n-2k)x],
   \sum_{k=0}^n\frac{(q\,;q)_n}{(q\,;q)_k(q\,;q)_{n-k}}\,e^{i(n-2k)x},
   \nonumber\\
   &\qquad\qquad H_0=1,\quad H_1(\cos x|q)=2\cos x.
\end{align}
It has a definite parity.
Reflecting the orthogonality of the eigenfunctions of the Hamiltonian
$\mathcal{H}$, $(\phi_n,\phi_m)\!\propto\delta_{nm}$,
it is orthogonal with respect to the weight function $\phi_0(x)^2$:
\begin{equation}
   \int_0^{\pi}\phi_0(x)^2H_n(\cos x|q)H_m(\cos x|q)dx
   =\delta_{n\,m}\frac{2\pi}{(q^{n+1};q)_\infty},
   \label{intphinphim}
\end{equation}
satisfying the three term recurrence relation
\begin{equation}
   2\eta H_n(\eta|q)=H_{n+1}(\eta|q)+(1-q^n)H_{n-1}(\eta|q).
\end{equation}

The action of the creation  $\mathcal{A}^\dagger$ and annihilation
$\mathcal{A}$ operators on the polynomial $H_n(\cos x|q)$ is
\begin{align}
   &\widetilde{\mathcal{A}}^{\dagger}\eqdef \phi_0^{-1}\circ
   \mathcal{A}^\dagger\circ \phi_0,\quad
   \widetilde{\mathcal{A}}\eqdef \phi_0^{-1}\circ
   \mathcal{A}\circ \phi_0,\\
   &\widetilde{\mathcal{A}}^{\dagger}=q^{-\frac12}\frac{-1}{z-z^{-1}}
   \bigl(z^{-2}q^{D/2}-z^2q^{-D/2}\bigr),\\
   &\widetilde{\mathcal{A}}
   =\frac{-1}{z-z^{-1}}\bigl(q^{D/2}-q^{-D/2}\bigr),
   \label{tildeA}\\
   &\widetilde{\mathcal{A}}^\dagger H_n(\cos x|q)
   =q^{-(n+1)/2}H_{n+1}(\cos x|q),\\
   &(\widetilde{\mathcal{A}}^\dagger)^n 1
   =q^{-n(n+1)/4}H_{n}(\cos x|q),\\
   &\widetilde{\mathcal{A}} H_n(\cos x|q)
   =(q^{-n/2}-q^{n/2})H_{n-1}(\cos x|q).
\end{align}
The similarity transformed $\widetilde{\mathcal{A}}$ \eqref{tildeA} is
proportional to the divided difference operator.

%%%%%%%%%%%%%%%%%%%%%%%%%%%%%%%%%%%%%%%%%%%%%%%%%%%%%%%%%%%%%%%
%                                                             %
%  4. Heisenberg operator solution                            %
%                                                             %
%%%%%%%%%%%%%%%%%%%%%%%%%%%%%%%%%%%%%%%%%%%%%%%%%%%%%%%%%%%%%%%
\section{Heisenberg operator solution}
\label{Heisenberg}

The harmonic oscillator is a typical example for which the Heisenberg
operator solution is known and the annihilation/creation operators can
also be extracted as the positive/negative frequency parts of the
Heisenberg operator solution.
The situation is parallel but slightly different for the $q$-oscillator.
The exact Heisenberg operator solution is derived and its
positive/negative frequency parts give another set of
annihilation/creation operators $a^{(\pm)}$ which are closely related to
$\mathcal{A}$ and $\mathcal{A}^\dagger$.
(For the general theory of exact Heisenberg operator solutions, see
\cite{os7,os12} for systems of  single degree of freedom and \cite{os9}
for a class of multi-particle dynamics.)

We start from the {\em closure relation\/}
\begin{align}
   [\mathcal{H},[\mathcal{H},\cos x]\,]&=\cos x\,R_0(\mathcal{H})
   +[\mathcal{H},\cos x]\,R_1(\mathcal{H}),
   \label{twocom}\\
   R_0(\mathcal{H})&\eqdef(q^{-\frac12}-q^{\frac12})^2(\mathcal{H}+1)^2,\\
   R_1(\mathcal{H})&\eqdef(q^{-\frac12}-q^{\frac12})^2(\mathcal{H}+1),
\end{align}
which can be readily verified.
This relation enables us to express any multiple commutator
\[
   [\mathcal{H},[\mathcal{H},\cdots,[\mathcal{H},\cos x]
   \!\cdot\!\cdot\cdot]]
\]
as a linear combination of the operators $\cos x$ and
$[\mathcal{H},\cos x]$ with coefficients depending on the Hamiltonian
$\mathcal{H}$ only.
Thus we arrive at the exact Heisenberg operator solution for the
{\em sinusoidal coordinate\/}  $\eta(x)\eqdef \cos x$ \cite{os7}:
\begin{align}
   e^{it\mathcal{H}}\cos x\,e^{-it\mathcal{H}}
   &=\cos x\,\frac{q\,e^{i\alpha_+(\mathcal{H})t}
   +e^{i\alpha_-(\mathcal{H})t}}{1+q}
   +\,[\mathcal{H},\cos x]\,
   \frac{e^{i\alpha_+(\mathcal{H})t}-e^{i\alpha_-(\mathcal{H})t}}
   {(q^{-1}-q){(\mathcal{H}+1)}},
    \label{quantsol}\\
   \alpha_{\pm}(\mathcal{H})&=(q^{\mp 1}-1)(\mathcal{H}+1).
\end{align}
This simply means that the coordinate $\cos x$ undergoes sinusoidal
motions with frequencies $\alpha_\pm({\mathcal H})$.

While factorization of Hamiltonian is known to provide the
annihilation/creation operators only for the harmonic oscillator and the
$q$-oscillator, the authentic definition of the annihilation/creation
operators is through the positive/negative frequency parts of the
Heisenberg operator solution \cite{os7} ($\eta=\cos x$):
\begin{align}
   &e^{it\mathcal{H}}\cos x\,e^{-it\mathcal{H}}
   =a^{(+)}\,e^{i\alpha_+(\mathcal{H})t}
   +a^{(-)}\,e^{i\alpha_-(\mathcal{H})t},
   \label{acdefs0}\\
   &a^{(\pm)}
   =\frac{\pm 1}{q^{-1}-q}\Bigl([\mathcal{H},\eta]_{q^{\pm 1}}
   +(1-q^{\pm 1})\eta\Bigr)
   (\mathcal{H}+1)^{-1},\nonumber\\
   &{a^{(-)}}^\dagger=a^{(+)}.
\end{align}
Their action on the full eigenfunction is
($\phi_n(x)\eqdef \phi_0(x)H_n(\cos x|q)$):
\begin{equation}
   a^{(-)}\phi_n=\tfrac12(1-q^n)\phi_{n-1},\quad
   a^{(+)}\phi_n=\tfrac12\phi_{n+1},
\end{equation}
to be compared with
\begin{equation}
   \mathcal{A}\phi_n=q^{-\frac{n}{2}}(1-q^n)\phi_{n-1},\quad
   \mathcal{A}^{\dagger}\phi_n=q^{-\frac{n+1}{2}}\phi_{n+1}.
\end{equation}
{}From these and \eqref{intphinphim}, it is easy to check the
hermiticity
\begin{align}
   (\phi_{n-1},a^{(-)}\phi_n)&=(a^{(+)}\phi_{n-1},\phi_n),\\
   (\phi_{n-1},\mathcal{A}\phi_n)&=
   (\mathcal{A}^{\dagger}\phi_{n-1},\phi_n).
\end{align}

They satisfy commutation relations
\begin{align}
   [a^{(-)},a^{(+)}]
   &=\tfrac14(1-q)(\mathcal{H}+1)^{-1},
   \label{apmcomm}\\
   [\mathcal{H},a^{(\pm)}]&=(q^{\mp 1}-1)a^{(\pm)}(\mathcal{H}+1).
   \label{KS3.26[H,apm]}
\end{align}
By removing the Hamiltonian from the r.h.s. they can be cast into
another $q$-oscillator form
\begin{align}
   a^{(-)}a^{(+)}-qa^{(+)}a^{(-)}&=\tfrac14(1-q),
   \label{secqosci}\\
   \mathcal{H}a^{(\pm)}-q^{\mp 1}a^{(\pm)}\mathcal{H}
   &=(q^{\mp 1}-1)a^{(\pm)}.
   \label{secqosci2}
\end{align}
It should be noted that  the $q$-oscillator relations
\eqref{secqosci}-\eqref{secqosci2} also hold for the continuous big
$q$-Hermite polynomial \cite{koeswart,os7}.
We will report on this topic elsewhere.

The two types of creation-annihilation operators are closely related
with each other \cite{os7}
\begin{equation}
   a^{(+)}=\mathcal{A}^\dagger X,\quad
   a^{(-)}=X^\dagger\mathcal{A},
\end{equation}
with
\begin{align}
   X&=-\frac{i}{2}q\bigl(z\sqrt{V(x)}\,q^{D/2}
   -z^{-1}\sqrt{V(x)^*}\,q^{-D/2}\,\bigr)
   (\mathcal{H}+1)^{-1},\\
   X^{\dagger}&=\frac{i}{2}q(\mathcal{H}+1)^{-1}
   \bigl(q^{D/2}z^{-1}\sqrt{V(x)^*}-q^{-D/2}z\sqrt{V(x)}\,\bigr),
\end{align}
and the operators $X$ and $X^\dagger$ map the eigenfunction $\phi_n$ to
itself:
\begin{equation}
   X\phi_n=\frac{1}{2}q^{(n+1)/2}\phi_n,\quad
   X^{\dagger}\phi_n=\frac{1}{2}q^{(n+1)/2}\phi_n.
\end{equation}
The structure of these operators is better understood by the similarity
transformation in terms of the groundstate wavefunction $\phi_0$
\begin{equation}
   \widetilde{X}\eqdef \phi_0^{-1}\circ X\circ \phi_0,\quad
   \widetilde{X}^{\dagger}\eqdef \phi_0^{-1}\circ X^\dagger\circ \phi_0.
\end{equation}
In fact, their actions on polynomials $\{H_n(\cos x|q)\}$ are essentially
identical:
\begin{align}
   \widetilde{X}&=\frac12q^{\frac12}\frac{-1}{z-z^{-1}}
   (z^{-1}q^{D/2}-zq^{-D/2})(\widetilde{\mathcal{H}}+1)^{-1},\\
   \widetilde{X}^{\dagger}&=\frac12q^{\frac12}
   (\widetilde{\mathcal{H}}+1)^{-1}\frac{-1}{z-z^{-1}}
   (z^{-1}q^{D/2}-zq^{-D/2}).
\end{align}
The main part of $\widetilde{X}$ and $\widetilde{X}^{\dagger}$,
defined by
\begin{equation}
   \mathcal{D}^q=\frac{-1}{z-z^{-1}}(z^{-1}q^{D/2}-z\,q^{-D/2}),
\end{equation}
was also introduced by Atakishiyev-Klimyk \cite{ataki2} eq(9).
It satisfies the relation
\begin{equation}
   \mathcal{D}^q H_n(\cos x|q)=q^{-n/2}H_n(\cos x|q),
\end{equation}
and it factorizes $\widetilde{\mathcal{H}}$ and
$\widetilde{\mathcal{H}}+1$:
\begin{equation}
   (\mathcal{D}^q-1)(\mathcal{D}^q+1)=\widetilde{\mathcal{H}},\quad
   (\mathcal{D}^q)^2=\widetilde{\mathcal{H}}+1.
\end{equation}

The coherent state of the harmonic oscillator is defined as the
eigenvector of the annihilation operator; $a\psi=\alpha\psi$,
which is the generating function of the Hermite polynomials.
We encounter a parallel situation here.
The eigenvector of the operator $a^{(-)}$,
$a^{(-)}\psi(x;\alpha)=\alpha\psi(x;\alpha)$, is given by
\begin{align}
   \psi(x\,;\alpha)&=\phi_0(x)\sum_{n=0}^\infty
   \frac{(2\alpha)^n}{(q\,;q)_n}H_n(\cos x|q)\\
   &=\phi_0(x)\frac{1}{(2\alpha\,e^{ix};q)_{\infty}
   (2\alpha\,e^{-ix};q)_{\infty}}.
\end{align}
The second factor is the generating function of the $q$-Hermite
polynomials \cite{koeswart,andaskroy}.
The coherent state defined by the other annihilation operator $\mathcal{A}$,
$\mathcal{A}\psi'(x;\alpha)=\alpha\psi'(x;\alpha)$, has a similar
structure:
\begin{equation}
   \psi'(x\,;\alpha)=\phi_0(x)\sum_{n=0}^\infty
   \frac{\alpha^nq^{\frac14n(n+1)}}{(q\,;q)_n}H_n(\cos x|q).
\end{equation}

%%%%%%%%%%%%%%%%%%%%%%%%%%%%%%%%%%%%%%%%%%%%%%%%%%%%%%%%%%%%%%%
%                                                             %
%  5. Limit to the ordinary harmonic oscillator               %
%                                                             %
%%%%%%%%%%%%%%%%%%%%%%%%%%%%%%%%%%%%%%%%%%%%%%%%%%%%%%%%%%%%%%%
\section{Limit to the ordinary harmonic oscillator}
\label{limit}

The $q$-oscillators reduce to the ordinary harmonic oscillator in the 
$q\to1$ limit.
To show this, let us introduce two parameters ($L$ and $c$) and a new
coordinate $x'$:
\begin{equation}
  x=\frac{\pi}{2}-\frac{\pi}{L}x'
  \ \ \Bigl(\Rightarrow\ -\frac{L}{2}<x'<\frac{L}{2}\Bigr),\quad
  q=e^{-\frac{2\pi}{cL}}.
\end{equation}
The momentum operator conjugate to $x'$ is
$p'=-i\frac{d}{dx'}=-\frac{\pi}{L}p$.
Then the desired limit is obtained by setting $L=\pi c$ and taking
$c\to\infty$ limit:
\begin{align}
  &c^2\mathcal{H}\to x^{\prime\,2}+p^{\prime\,2}-1,\quad
  c^2\mathcal{E}_n\to 2n,\\
  &\genfrac{.}{\}}{0pt}{}{c\mathcal{A}^{\dagger}}{c\mathcal{A}}
  \to x'\mp ip',\quad
  ca^{(\pm)}\to\frac12(x'\mp ip'),\quad
  \genfrac{.}{\}}{0pt}{}{X^{\dagger}}{X}\to\frac12\,,\\
  &c\cos x\to x'\ \ (-\infty<x'<\infty),\quad
  c^4R_0(\mathcal{H})\to 4,\quad
  c^2R_1(\mathcal{H})\to 0,\\
  &\frac{(q\,;q)_{\infty}\phi_0(x)^2}{2\sqrt{\pi}\,c}
  \to e^{-x^{\prime\,2}},
  \label{phi0limit}\\
  &c^nH_n(\cos x|q)=c^nH_n\bigl(\sin\frac{x'}{c}\bigm|e^{-\frac{2}{c^2}}\bigr)
  \to H_n(x').
  \label{H_nlimit}
\end{align}
Here we have used the Jacobi's triple product identity \cite{andaskroy} 
and its modular transformation property (the $S$-transformation)
for deriving \eqref{phi0limit},
and the three term recurrence relations for \eqref{H_nlimit}.

%%%%%%%%%%%%%%%%%%%%%%%%%%%%%%%%%%%%%%%%%%%%%%%%%%%%%%%%%%%%%%%
%                                                             %
%  6. Other forms of $q$-oscillators                          %
%                                                             %
%%%%%%%%%%%%%%%%%%%%%%%%%%%%%%%%%%%%%%%%%%%%%%%%%%%%%%%%%%%%%%%
\section{Other forms of $q$-oscillators}
\label{other}

Here we will discuss the relationship between our intrinsic
$q$-oscillator algebra \eqref{qrel}-\eqref{qrel2} and those introduced
purely algebraically for quantum group representations around 1989-90
\cite{qoscillators}.
First let us introduce the number operator $\mathcal{N}$ through the
energy spectrum formula \eqref{spectrum},
\begin{equation}
   (\mathcal{H}+1)^{\mp1}=q^{\pm\mathcal{N}},\quad
   \mathcal{N}\phi_n=n\phi_n,\quad n\in\mathbb{Z}_+,
\end{equation}
which counts the level from the groundstate.
Several different forms of $q$-oscillator algebras are introduced,
among which we list two typical ones:
\begin{align}
   bb^{\dagger}-q^{-1}b^{\dagger}b&=q^{\mathcal{N}},
   \label{qosci1}\\
   bb^{\dagger}-q\,b^{\dagger}b&=q^{-\mathcal{N}}.
   \label{qosci2}
\end{align}
If we define $b$ and $b^{\dagger}$ by
\begin{align}
   b=\frac{\mathcal{A}\,q^{\mathcal{N}/4}}
   {(q^{-\frac{1}{2}}-q^{\frac{1}{2}})^{\frac{1}{2}}},\quad
   b^{\dagger}=\frac{q^{\mathcal{N}/4}\mathcal{A}^{\dagger}}
   {(q^{-\frac{1}{2}}-q^{\frac{1}{2}})^{\frac{1}{2}}},
\end{align}
it is straightforward to verify
\begin{equation}
   bb^{\dagger}-q^{-\frac12}b^{\dagger}b=q^{\mathcal{N}/2},
\end{equation}
which becomes \eqref{qosci1} by identification $q\to q^2$.
Likewise, the $q$-oscillator algebra of $a^{(\pm)}$ \eqref{secqosci} is
related to \eqref{qosci2} by similar transformations.

%%%%%%%%%%%%%%%%%%%%%%%%%%%%%%%%%%%%%%%%%%%%%%%%%%%%%%%%%%%%%%%
%                                                             %
%  7. Comments and summary                                    %
%                                                             %
%%%%%%%%%%%%%%%%%%%%%%%%%%%%%%%%%%%%%%%%%%%%%%%%%%%%%%%%%%%%%%%
\section{Comments and summary}
\label{comments}

Some historical comments are in order.
There were attempts to relate $q$-oscillator algebras to the difference
equation of the $q$-Hermite polynomial. None of them is based on a
Hamiltonian, thus hermiticity is not manifest and the logic for
factorization is unclear.
Here we list a few such attempts.
Atakishiyev and Suslov in 1990 \cite{ataki1} wrote down an algebra
\begin{equation}
   bb^+-q^{-1}b^+b=1,\quad H=b^+b,
\end{equation}
which is related to our $q$-oscillator algebra \eqref{qrel} by a
similarity transformation
\begin{equation}
   \sqrt{q^{-1}-1}\,\genfrac{(}{)}{0pt}{}{b}{b^+}
   =\frac{1}{\sqrt{\sin x}}\circ
   \genfrac{(}{)}{0pt}{}{\mathcal{A}}{\mathcal{A}^\dagger}
   \circ \sqrt{\sin x}.
\end{equation}
Floreanini, LeTourneux and Vinet presented in 1994 \cite{floreanini}
a $q$-oscillator algebra
($\widetilde{\mathcal N}\eqdef \phi_0^{-1}\circ\mathcal{N}\circ\phi_0$)
\begin{equation}
   A_-A_+-q^{-1}A_+A_-=1,\quad
   KA_\pm=q^{\mp\frac{1}{2}}A_\pm K,\quad
   K=q^{-\widetilde{\mathcal N}/2},
\end{equation}
which is in our notation
\begin{equation}
   A_+=-\widetilde{\mathcal{A}}^{\dagger},\quad
   A_-=\frac{-1}{q^{-1}-1}\,\widetilde{\mathcal{A}},\quad
   K=\mathcal{D}^q.
\end{equation}
In 2003 Borzov and Damaskinsky \cite{qoscillators2} wrote down
\begin{equation}
   a_q^-a_q^+-q\,a_q^+a_q^-=1,
\end{equation}
starting from the three term recurrence relation of the $q$-Hermite
polynomial and defining the annihilation/\hspace{0pt}creation operators
in their own way.

In summary: we have derived two $q$-oscillator algebras \eqref{qrel} and
\eqref{secqosci} from the Hamiltonian of the $q$-Hermite polynomial
\eqref{factorise}--\eqref{defA} \cite{os5,os7,os12}, which is a special
case of the Askey-Wilson polynomial \cite{koeswart,andaskroy}.
The generators are
genuine annihilation/creation operators and the hermiticity is manifest.

%\bigskip
%%%%%%%%%%%%%%%%%%%%%%%%%%%%%%%%%%%%%%%%%%%%%%%%%%%%%%%%%%%%%%%
%                                                             %
%  Acknowledgments                                            %
%                                                             %
%%%%%%%%%%%%%%%%%%%%%%%%%%%%%%%%%%%%%%%%%%%%%%%%%%%%%%%%%%%%%%%
\section*{Acknowledgments}
This work is supported in part by Grants-in-Aid for Scientific Research
from the Ministry of Education, Culture, Sports, Science and Technology,
No.18340061 and No.19540179.

%%%%%%%%%%%%%%%%%%%%%%%%%%%%%%%%%%%%%%%%%%%%%%%%%%%%%%%%%%%%%%%
%                                                             %
%  References                                                 %
%                                                             %
%%%%%%%%%%%%%%%%%%%%%%%%%%%%%%%%%%%%%%%%%%%%%%%%%%%%%%%%%%%%%%%

\end{document}